\documentclass[%
preprint,
tighten,
amsmath,
amssymb,
aps,
twocolumn,
]{aastex631}

\usepackage{graphicx}
\usepackage{dcolumn}
\usepackage{bm}
\usepackage{color}
\usepackage{natbib}
\usepackage{comment}
\newcommand{\gal}{UGC 2698}
\newcommand{\overs}{4}

\newcommand{\ds}{4}
\newcommand{\msig}{$M_{\mathrm{BH}}-\sigma_\star$}
\newcommand{\mlum}{$M_{\mathrm{BH}}-L_{\mathrm{bul}}$}
\newcommand{\mmass}{$M_{\mathrm{BH}}-M_{\mathrm{bul}}$}

\newcommand{\mbh}{$M_{\mathrm{BH}}$}
\newcommand{\inc}{$i$}
\newcommand{\ml}{$M/L_\mathrm{H}$}
\newcommand{\pa}{$\Gamma$}
\newcommand{\vsys}{$v_{\text{sys}}$}
\newcommand{\xloc}{$x_0$}
\newcommand{\yloc}{$y_0$}
\newcommand{\fw}{$f_0$}
\newcommand{\sig}{$\sigma_0$}

\newcommand{\kms}{km s$^{-1}$}

\newcommand{\mbhfullerr}{\mbh\ $= (2.46 \pm{0.07}$ [$1\sigma$ stat] $^{+0.70}_{-0.78}$ [sys])$\times 10^9$ $M_\odot$}
\newcommand{\mbhfullerrforabstract}{\mbh\ $= (2.46\pm{0.07}$ [$1\sigma$ statistical] $^{+0.70}_{-0.78}$ [systematic])$\times 10^9$ $M_\odot$}

\begin{document}

\shortauthors{Cohn et al.}

\title{An ALMA Gas-dynamical Mass Measurement of the Supermassive Black Hole in the Local Compact Galaxy \gal}

\author[0000-0003-1420-6037]{Jonathan H. Cohn}
\affil{George P. and Cynthia W. Mitchell Institute for Fundamental Physics and Astronomy, Department of Physics \& Astronomy, Texas A\&M University, 4242 TAMU, College Station, TX 77843, USA}

\author[0000-0002-1881-5908]{Jonelle L. Walsh}
\affiliation{George P. and Cynthia W. Mitchell Institute for Fundamental Physics and Astronomy, Department of Physics \& Astronomy, Texas A\&M University, 4242 TAMU, College Station, TX 77843, USA}

\author[0000-0001-6301-570X]{Benjamin D. Boizelle}
\affiliation{Department of Physics and Astronomy, N284 ESC, Brigham Young University, Provo, UT, 84602, USA}
\affiliation{George P. and Cynthia W. Mitchell Institute for Fundamental Physics and Astronomy, Department of Physics \& Astronomy, Texas A\&M University, 4242 TAMU, College Station, TX 77843, USA}

\author[0000-0002-3026-0562]{Aaron J. Barth}
\affiliation{Department of Physics and Astronomy, 4129 Frederick Reines Hall, University of California, Irvine, CA, 92697-4575, USA}

\author[0000-0002-8433-8185]{Karl Gebhardt}
\affiliation{Department of Astronomy, The University of Texas at Austin, 2515 Speedway, Stop C1400, Austin, TX 78712, USA}

\author[0000-0002-1146-0198]{Kayhan G\"{u}ltekin}
\affiliation{Department of Astronomy, University of Michigan, 1085 S. University Ave., Ann Arbor, MI 48109, USA}

\author[0000-0003-1693-7669]{Ak\i n Y\i ld\i r\i m}
\affiliation{Max-Planck-Institut f\"{u}r Astrophysik, Karl-Schwarzschild-Str. 1, 85748 Garching, Germany}

\author[0000-0002-3202-9487]{David A. Buote}
\affiliation{Department of Physics and Astronomy, 4129 Frederick Reines Hall, University of California, Irvine, CA, 92697-4575, USA}

\author[0000-0003-2511-2060]{Jeremy Darling}
\affiliation{Center for Astrophysics and Space Astronomy, Department of Astrophysical and Planetary Sciences, University of Colorado, 389 UCB, Boulder, CO 80309-0389, USA}

\author[0000-0002-7892-396X]{Andrew J. Baker}
\affiliation{Department of Physics and Astronomy, Rutgers, the State University of New Jersey, 136 Frelinghuysen Road Piscataway, NJ 08854-8019, USA}

\author[0000-0001-6947-5846]{Luis C. Ho}
\affiliation{Kavli Institute for Astronomy and Astrophysics, Peking University, Beijing 100871, China; Department of Astronomy, School of Physics, Peking University, Beijing 100871, China}

\author[0000-0003-2632-8875]{Kyle M. Kabasares}
\affiliation{Department of Physics and Astronomy, 4129 Frederick Reines Hall, University of California, Irvine, CA, 92697-4575, USA}

\correspondingauthor{Jonathan H. Cohn}
\email{joncohn@tamu.edu}

\begin{abstract}

We present 0\farcs{14}-resolution Atacama Large Millimeter/submillimeter Array (ALMA) CO(2$-$1) observations of the circumnuclear gas disk in \gal, a local compact galaxy.
The disk exhibits regular rotation with projected velocities rising to $450$ \kms\ near the galaxy center. We fit gas-dynamical models to the ALMA data cube, assuming the CO emission originates from a dynamically cold, thin disk, and measured the mass of the supermassive black hole (BH) in \gal\ to be \mbhfullerrforabstract.
\gal\ is part of a sample of nearby early-type galaxies that are plausible $z\sim2$ red nugget relics.
Previous stellar-dynamical modeling for three galaxies in the sample found BH masses consistent with the BH mass$-$stellar velocity dispersion (\msig) relation but over-massive relative to the BH mass$-$bulge luminosity (\mlum) correlation, suggesting that BHs may gain the majority of their mass before their host galaxies.
However, \gal\ is consistent with both \msig\ and \mlum.
As \gal\ has the largest stellar mass and effective radius in the local compact galaxy sample, it may have undergone more recent mergers that brought it in line with the BH scaling relations.
Alternatively, given that the three previously-measured compact galaxies are outliers from \mlum, while \gal\ is not, there may be significant scatter at the poorly sampled high-mass end of the relation.
Additional gas-dynamical \mbh\ measurements for the compact galaxy sample will improve our understanding of BH$-$galaxy co-evolution.
\end{abstract}

\section{\label{intro}Introduction}

Supermassive black holes (BHs) are found at the centers of most or all massive galaxies \citep{Magorrian1998,Kormendy2013}.
Furthermore, BH mass (\mbh) correlates with large-scale properties of these galaxies, such as bulge luminosity ($L_\mathrm{bul}$), bulge mass ($M_\mathrm{bul}$), and stellar velocity dispersion ($\sigma_\star$), despite BHs only dominating the gravitational potential in the central regions of galaxies (e.g., \citealt{Kormendy1995,Ferrarese2000,Gebhardt2000,Marconi2003,Gultekin2009,Kormendy2013,Saglia2016}).
These correlations indicate that BHs co-evolve with their host galaxies, but the extreme high- and low-mass ends of the scaling relations are poorly sampled, and it remains uncertain whether the relations apply to every type of galaxy (e.g., \citealt{Lauer2007,Greene2010,McConnell2012,Kormendy2013,McConnellMa2013,Rusli2013a,Greene2016,Lasker2016,Thomas2016,Greene2020}).
As such, the exact details of how BHs and galaxies grow and evolve together over time are not well understood.

Currently, there are $\sim$100 dynamical BH detections \citep{Saglia2016}.
One of the main methods for measuring \mbh\ involves modeling the rotation of circumnuclear gas disks.
Traditionally, these gas-dynamical models have fit to Hubble Space Telescope (HST) observations of ionized gas (e.g., \citealt{Macchetto1997, Barth2001}), but molecular gas, which tends to exhibit less turbulent motion (e.g., \citealt{Alatalo2013,Barth2016a,Boizelle2019}), offers an attractive alternative.
The first molecular gas-dynamical \mbh\ measurement was made by \citet{Davis2013}, who used the Combined Array for Research in Millimeter-wave Astronomy to study the early-type galaxy (ETG) NGC 4526.
Even though NGC 4526 is nearby ($D=16.4$ Mpc), the authors required several hours on-source to detect nuclear CO(2$-$1) emission and marginally resolve the BH sphere of influence (SOI; $r_{\mathrm{SOI}} = G$\mbh$/\sigma_\star^2$), where the BH dominates the galaxy's gravitational potential.
Nevertheless, the work set a new precedent for measuring BH masses.

Since then, the Atacama Large Millimeter/sub-millimeter Array (ALMA) has revolutionized gas-dynamical BH measurements.
With significantly improved angular resolution and sensitivity compared to previous mm/sub-mm interferometers, ALMA is able to efficiently detect molecular gas within $r_{\mathrm{SOI}}$ for galaxies at larger distances ($\sim$100 Mpc).
As a result, there has been a large increase in the number of gas-dynamical \mbh\ measurements (e.g., \citealt{Barth2016a,Davis2017,Onishi2017,Davis2018,Boizelle2019,Nagai2019,North2019,Smith2019,Davis2020,Nguyen2020,Boizelle2021}).

Recently, the Hobby-Eberly Telescope Massive Galaxy Survey (HETMGS) obtained optical long-slit spectra for over 1000 nearby galaxies selected to cover various combinations of stellar mass, size, and stellar velocity dispersion \citep{Bosch2015}.
One intriguing result from HETMGS was the identification of an unusual sample of local compact galaxies.
Based on the \msig\ relation \citep{Kormendy2013, Saglia2016}, these galaxies are expected to have large BHs with \mbh\ up to $\sim6\times10^9\ M_{\odot}$.
However, the galaxies that usually host such massive BHs in the local universe are giant elliptical galaxies and brightest cluster galaxies (BCGs), which tend to be round and dispersion-supported, with large effective radii ($\sim$10 kpc) and cored central surface brightness profiles (e.g., \citealt{Bonta2009, McConnell2011a, McConnell2011b, McConnell2012, Thomas2016, Mehrgan2019}). 

Instead of looking like nearby giant ellipticals and BCGs, the local compact galaxies from HETMGS resemble $z\sim2$ massive, quiescent galaxies (red nuggets; e.g., \citealt{Daddi2005,Trujillo2007,Dokkum2008}).
There are 15 local compact galaxies from HETMGS identified in \cite{Yildirim2017}, and they have small effective radii ($0.7-3.1$ kpc) for their large stellar masses ($5.5\times10^{10} - 3.8\times10^{11}\ M_\odot$).
These galaxies are consistent with the $z\sim2$ mass$-$size relation instead of the local relation.
They also have cuspy surface brightness profiles and are flattened and rotating.
In some cases, the HETMGS compact galaxies also have uniformly old stellar ages ($\gtrsim$10 Gyr; \citealt{Martin2015,Mateu2017}) over several effective radii, stellar orbital distributions that do not show evidence for major mergers \citep{Yildirim2017}, high dark matter halo concentrations \citep{Buote2018,Buote2019}, and associated globular cluster systems with red color distributions \citep{Beasley2018}.
Therefore, it has been suggested that the local compact galaxies may be passively-evolved relics of the $z\sim2$ red nuggets (e.g.,  \citealt{Trujillo2014,Mateu2015,Yildirim2017}).

There are already stellar-dynamical \mbh\ determinations for three objects in the compact galaxy sample: NGC 1277, NGC 1271, and Mrk 1216 (e.g., \citealt{Bosch2012,Emsellem2013,Walsh2015,Yildirim2015,Graham2016a,Walsh2016,Walsh2017,Krajnovic2018}).
These galaxies are consistent with the \msig\ relation and are over-massive relative to the \mlum\ and \mmass\ scaling relations (\citealt{Walsh2015,Walsh2016,Walsh2017}), although the magnitude of the offset depends on assumptions about the bulge luminosity/mass (e.g., \citealt{SavorgnanGraham2016,Graham2016b}).
If the galaxies are red nugget relics and contain over-massive BHs, that could indicate that BHs tend to acquire the majority of their mass by $z\sim2$, after which most massive elliptical galaxies grow via minor-intermediate dry mergers that do not add much mass to their BHs (e.g., \citealt{Dokkum2010,Hilz2013}).
Obtaining more \mbh\ measurements for the HETMGS compact galaxy sample will help further populate the high-mass end of the scaling relations.
This sample also allows us to explore galaxies with growth histories that are very different from those of the typical hosts of such massive BHs.

In this paper, we analyze \gal, one of the local compact galaxies that does not yet have an \mbh\ measurement.
\gal\ has the largest stellar mass ($3.8\times10^{11}\ M_\odot$) and effective radius (3.1 kpc) of the sample, with a stellar velocity dispersion $\sigma_\star = 304$ km s$^{-1}$ \citep{Yildirim2017}.
The galaxy hosts a regular nuclear dust disk visible in an HST $I$-band image, which suggests the presence of cleanly rotating molecular gas (e.g., \citealt{Alatalo2013}).
We obtained 0\farcs{14}-resolution ALMA Cycle 4 CO(2$-$1) observations that allowed us to investigate the spatial and kinematic structure of the molecular gas disk.
By constructing dynamical models that fit directly to the ALMA data cube, we derive a molecular gas-dynamical BH mass.

The paper is structured as follows.
In \S\ref{observations} we present the HST and ALMA data.
We decompose the stellar light distribution with Multi-Gaussian Expansions and discuss the nuclear CO(2$-$1) emission in \gal.
In \S\ref{model}, we describe our dynamical model and optimization method.
In \S\ref{results}, we present the results of the modeling, and in \S\ref{discussion}, we estimate the \gal\ BH SOI, compare to other local compact galaxies with dynamical \mbh\ measurements, and discuss implications for BH$-$galaxy co-evolution.
We summarize our findings in \S\ref{conclusions}.
Throughout the paper, we assume an angular diameter distance of 91 Mpc, where 441.2 pc spans 1\arcsec.
This comes from a $\Lambda$CDM cosmology with H$_0 = 67.8$ km s$^{-1}$ Mpc$^{-1}$, $\Omega_M = 0.308$, and $\Omega_{\Lambda} = 0.692$, and adopting the Hubble Flow distance from the Virgo + Great Attractor + Shapley Supercluster infall model \citep{Mould2000} from the NASA/IPAC Extragalactic Database\footnote{\url{https://ned.ipac.caltech.edu/}}.
We note that the measured BH mass scales linearly with the assumed distance.

\section{\label{observations}Observations}

Constraining \mbh\ with molecular gas-dynamical models requires characterizing a galaxy's stellar light in order to account for the stellar contribution to the gravitational potential, in addition to high-resolution measurements of the gas near the center of the galaxy.
We discuss our resolution in more depth in \S\ref{resolution}.
Below, we describe the HST Wide Field Camera 3 (WFC3) observations and our parameterization of the stellar light profile.
We also detail the ALMA observations and discuss the structure and kinematics of the CO(2$-$1) emission in \gal.

\subsection{HST Data\label{hst}}
\gal\ was observed on 2013 August 28 under program GO-13416 (PI: van den Bosch) with the HST WFC3 IR F160W (\textit{H}-band) and UVIS F814W (\textit{I}-band) filters.
The $H$-band observations included dithered full-array exposures and a series of short sub-array exposures taken in a four-point dither pattern to better sample the point spread function (PSF) and to avoid saturating the nucleus.
The $I$-band observations consisted of three dithered full-array exposures.

\begin{figure*}
\includegraphics[width=\textwidth]{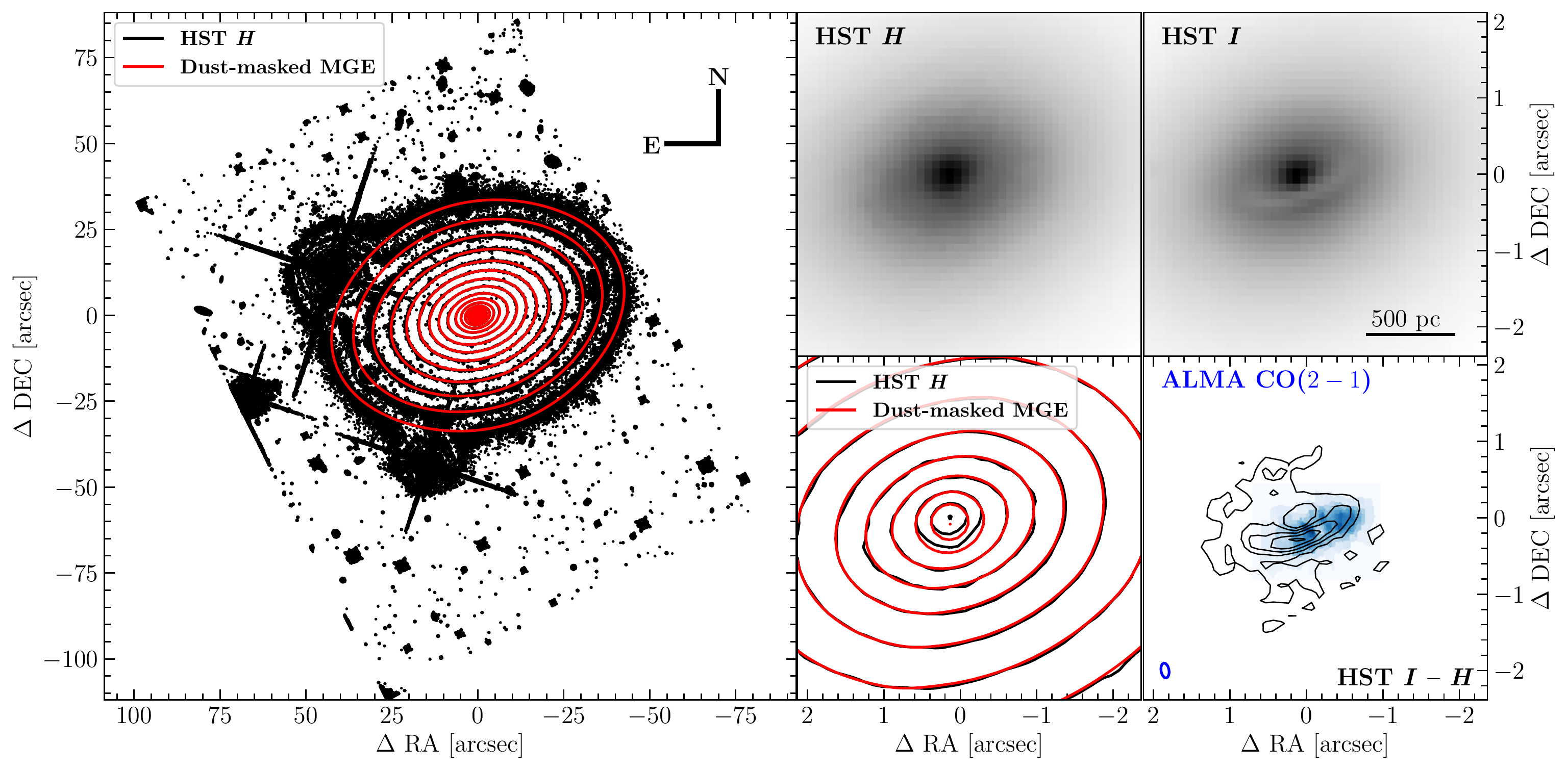}
\caption{(\textit{Left}) HST F160W ($H$-band) image of \gal\ (black contours) with the dust-masked MGE overlaid (red contours).
(\textit{Center, top}) Zoom-in of the inner 4\farcs{5} of the HST $H$-band image.
(\textit{Right, top}) Zoom-in of the inner 4\farcs{5} of the HST F814W ($I$-band) image.
A regular, inclined, $1\farcs{6}$-diameter dust disk is seen in the $I$-band image, but is not evident in the $H$-band image.
(\textit{Center, bottom}) Zoom-in of the inner 4\farcs{5} of the HST $H$-band image (black contours) and the dust-masked MGE (red contours).
There is a slight asymmetry in the black contours near the galaxy center, indicating the presence of dust.
The red contours correspond to the dust-masked MGE, which was generated by fitting to the $H$-band image with the dust-contaminated region masked out; in these contours, the asymmetry is no longer present.
(\textit{Right, bottom}) Zoom-in of the inner 4\farcs{5} of the HST $I-H$ image (black contours) overlaid on the ALMA CO(2$-$1) emission (blue).
The blue ellipse at the bottom left of the panel represents the synthesized ALMA beam.
The CO(2$-$1) emission is co-spatial with the dust disk and is approximately the same diameter.}
\label{fig_hst}
\end{figure*}

The data were processed with the {\tt calwf3} pipeline and then cleaned, distortion corrected, and combined with {\tt AstroDrizzle} \citep{Gonzaga2012}.
The resultant $H$-band image has a $0\farcs1$ pixel$^{-1}$ scale and covers a $2\farcm7\times2\farcm1$ field of view (FOV), with a total exposure time of 898.5 s.
Since we used the $I$-band image solely for the purposes of constructing an $I-H$ map, we drizzled the $I$-band observations onto the same 0\farcs1 pixel scale as the $H$-band image.
The total exposure time of the final $I$-band image is 805.0 s.
The HST $H$, $I$, and $I-H$ images are shown in Figure \ref{fig_hst}.

A circumnuclear dust disk or ring-like structure is visible in the $I$-band image, spanning $\sim$1\farcs{6} in width.
No such dust disk is immediately apparent in the $H$-band image.
In the $I-H$ map, we find a maximum color excess in the disk of 0.70 mag, located 0\farcs{33} to the south/southwest of the nucleus.
This excess is measured relative to the median $I-H$ color just outside the disk, which corresponds to 1.73 mag.

\subsubsection{Describing the Surface Brightness with Multi-Gaussian Expansion Models\label{mge}}

We described the galaxy's surface brightness using a Multi-Gaussian Expansion (MGE).
This description is not physically motivated, but MGEs are known to accurately reproduce the stellar profiles of ETGs \citep{Emsellem1994,Cappellari2002}.
We fit the sum of two-dimensional (2D) Gaussians to the $H$-band mosaic using {\tt GALFIT} \citep{Peng2010}, with results from the \cite{Cappellari2002} {\tt mgefit} package as the initial parameter guesses.
During the {\tt GALFIT} run, we accounted for the $H$-band PSF, which was constructed using Tiny Tim \citep{Krist2004} models dithered and drizzled in a fashion identical to the HST observations.
We also included a mask when running {\tt GALFIT} to exclude foreground stars, galaxies, and detector artifacts.

During the fit, the Gaussian components were constrained to have an identical position angle (PA) and centroid.
The background was included in {\tt GALFIT} as a fixed constant and was determined by calculating the mean of various regions in the south/southeast corner of the $H$-band image, which is the farthest corner from the galaxy nucleus and is located close to the galaxy minor axis.
The resulting $H$-band MGE (hereafter, the original MGE) has 11 Gaussian components, with projected axis ratios $q^\prime$ ranging from 0.538 to 1.000, projected dispersions $\sigma^\prime$ between 0\farcs{103} and 45\farcs{979}, and a PA of $108.755^\circ$ east of north.
The MGE is a good match to the image, with typical residuals $\sim$0.3$\%$ of the data.

While dust is clearly present in the HST \textit{I}-band data, the HST \textit{H}-band image is not as obviously affected by dust (see Figure \ref{fig_hst}).
However, upon close inspection, the inner contours of the $H$-band image are slightly asymmetric, with the surface brightness suppressed on the southern side of the nucleus compared to the northern side, indicating there is some dust extinction.
We therefore attempted to use two methods to account for dust: first, we generated an MGE for the $H$-band image after masking dust-contaminated regions, and then we constructed another MGE after applying a rough dust correction to the dust-masked $H$-band image.

To construct the dust mask, we started by flagging pixels that are redder than $I - H = 1.84$ mag.
For comparison, the average color is $\sim$1.73 mag just beyond the disk region.
We fit an MGE to the initial dust-masked $H$-band image with {\tt GALFIT}, following the same process used to construct the original MGE.
Based on our inspection of the {\tt GALFIT} residuals near the nucleus, we excluded additional pixels and re-fit the MGE.
This iterative process continued until the resultant {\tt GALFIT} residuals were $\lesssim$5\% and the mask contained the entire near side of the dust disk.
The MGE parameters found when fitting to the dust-masked $H$-band image (hereafter, the dust-masked MGE) are shown in Table \ref{tab_mge}, and the MGE is compared to the original $H$-band image in Figure \ref{fig_hst}.
This MGE is an excellent match to the dust-masked $H$-band image, with residuals at the $\sim$0.2$\%$ level.

\begin{deluxetable}{cccc}[ht]
\tabletypesize{\small}
\tablecaption{Dust-masked MGE parameters}
\tablewidth{0pt}
\tablehead{
\colhead{$j$} & 
\colhead{$\log_{10}(I_{H,j})$ [$L_{\odot}$ pc$^{-2}$]} & 
\colhead{$\sigma_j^\prime$ [arcsec]} & 
\colhead{$q_j^\prime$}
\\[-1.5ex]
\colhead{(1)} & 
\colhead{(2)} & 
\colhead{(3)} & 
\colhead{(4)}
}
\startdata
1 & 5.143 & 0.097 & 1.000 \\
2 & 4.633 & 0.217 & 1.000 \\
3 & 4.609 & 0.566 & 0.682 \\
4 & 4.065 & 1.043 & 0.740 \\
5 & 3.762 & 1.511 & 0.734 \\
6 & 3.526 & 2.661 & 0.750 \\
7 & 2.881 & 5.264 & 0.612 \\
8 & 2.919 & 7.868 & 0.720 \\
9 & 1.582 & 14.957 & 1.000 \\
10 & 2.216 & 17.001 & 0.653 \\
11 & 1.348 & 45.723 & 0.819 \\
\enddata
\begin{singlespace}
\tablecomments{MGE parameters found by fitting the dust-masked HST \textit{H}-band image of \gal.
Column (1) shows the MGE component number.
Column (2) gives the central surface brightness of each component, based on an absolute \textit{H}-band magnitude of 3.37 mag for the Sun \citep{Willmer2018} and a Galactic extinction of 0.075 mag \citep{Schlafly2011}.
Column (3) provides the projected dispersion of the Gaussian component along the major axis, and Column (4) lists the axis ratio.
The primed parameters are projected quantities. All components have a PA of $108.764^\circ$ east of north.}
\end{singlespace}
\label{tab_mge}
\end{deluxetable}

In addition to a dust-masked MGE, we applied a dust correction to the $H$-band image, following the methods outlined in \citet{Viaene2017} and \citet{Boizelle2019,Boizelle2021}.
As a brief summary, we assumed the inclined dust to be geometrically thin and embedded in the galaxy's midplane.
When comparing $I-H$ for pixels in the disk relative to the color outside of the disk, if the disk has low optical depth, then the observed color excess $\Delta(I-H)$ will be small and will scale approximately linearly with the intrinsic extinction.
As the opacity increases, the color excess will increase only to a turnover point, after which the color excess decreases to zero, as the light originating behind the disk becomes completely obscured while the light originating in front of the disk remains unaffected.
Moreover, the color excess should vary spatially across the disk for a given optical depth.
Along sightlines to the near side of the inclined disk, a larger fraction of light will come from behind the disk, and the color excess will therefore be larger than along sightlines to the far side of the disk.

\begin{figure}
\includegraphics[width=0.47\textwidth]{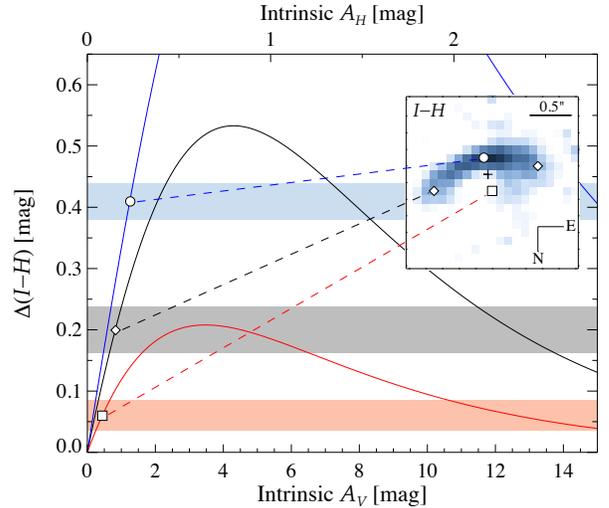}
\caption{Model color excess curves as a function of intrinsic extinction in the $V$-band (bottom x-axis) or $H$-band (top x-axis).
The model curves (red, gray, and blue) are presented for three example locations in the dust disk, denoted in the $I-H$ inset panel by the square, circle, and the diamond points.
The observed color excess at each example location is displayed as a shaded red, gray, or blue horizontal band.
The width of each horizontal band corresponds to the uncertainty in the observed color excess at the given point in the $I-H$ image. Along the major axis, the observed color excess is taken to be the average of the measured values on both the approaching and receding sides (diamond points).
Each observed color excess intersects its model curve twice -- once at a low intrinsic extinction value and once at a higher extinction.
As seen in Figure \ref{fig_hst}, there is no obvious dust disk in the HST $H$-band image.
Therefore, we assume the smaller of the two intrinsic $A_V$ values is the true extinction when correcting the image for dust.}
\label{fig_dust}
\end{figure}

We assume that \gal\ is oblate axisymmetric and that the gas disk, dust disk, and galaxy have the same inclination angle, taken to be $i = 68^\circ$ based on initial gas-dynamical modeling runs.
We deprojected the dust-masked MGE to obtain a three-dimensional model of the galaxy.
From this model, we calculated the fraction of stellar light as a function of spatial location that originates behind the embedded, inclined dust disk.
We determined the model color excess, $\Delta(I-H)$, as a function of intrinsic dust extinction ($A_V$) assuming a standard Galactic extinction law ($R_V = 3.1$; \citealt{Cardelli1989}).
The model color excesses at three example locations in the dust disk are shown in Figure \ref{fig_dust}.
As expected, the model color excess is not a monotonically increasing function of intrinsic extinction.
At the turnover point, the intersection of the observed color excess with $\Delta(I-H)$ yields a unique intrinsic $A_V$; however, below the turnover value, there exist two possible solutions (as was the case for NGC 3258; \citealt{Boizelle2019}).
As there is no obviously discernible dust in the $H$-band image, we assumed the exceptionally large extinction value ($A_V \gtrsim 10$ mag) is incorrect and adopted the lower intrinsic extinction value.
We found a maximum intrinsic extinction along the major axis between $A_V \sim 0.80 - 0.96$ mag ($A_H \sim 0.17 - 0.20$ mag), with the disk most opaque at a distance of $\sim$0$\farcs{52}$ along the minor axis.

With an estimate of the intrinsic extinction at each pixel in the disk, we calculated a dust-corrected $H$-band image.
We note that proper recovery of the intrinsic stellar surface brightness requires radiative transfer models that take into account the disk geometry, thickness, dust scattering, and extinction within the disk.
Nevertheless, this simple approach provides a way to gauge the possible impact of dust on the inferred \mbh.
We fit an MGE with {\tt GALFIT} to the dust-corrected $H$-band image.
This MGE (hereafter, the dust-corrected MGE) is composed of nine Gaussians, with $q^\prime$ ranging from 0.614 to 0.998 and $\sigma^\prime$ ranging from 0\farcs{042} to 47\farcs{112}.
The semi-major axis PA is $108.758^\circ$ east of north for all components.
The dust-corrected MGE has typical residuals of $\sim$0.3$\%$ relative to the data.

\begin{figure}
\includegraphics[width=0.47\textwidth]{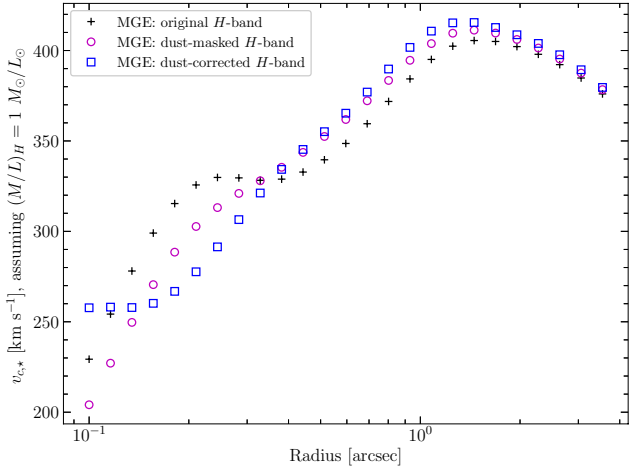}
\caption{Circular velocity as a function of radius resulting from each of the three \gal\ MGE models, assuming $M/L_H = 1\ M_\odot/L_\odot$.
Each MGE differs slightly in the inner $\sim$2\arcsec\ of the galaxy.
We use the dust-masked MGE (magenta circles) in the fiducial dynamical model.}
\label{vcirc}
\end{figure}

Ultimately, we have three MGEs -- one determined from a fit to the original $H$-band image (original MGE), one constructed from fitting a dust-masked $H$-band image (dust-masked MGE), and one found from matching to a dust-corrected $H$-band image (dust-corrected MGE).
In order to highlight differences in the three MGEs, in Figure \ref{vcirc} we show the resultant circular velocity due to the stars (based on each of the MGE models) as a function of radius, assuming an $H$-band stellar mass-to-light ratio ($M/L_H$) of $1$ $M_\odot/L_\odot$.
The MGEs produce slightly different velocity profiles in the central $\sim$2$\arcsec$ of the galaxy.
We use the dust-masked MGE in our fiducial dynamical model as it produces the best $\chi^2$, but we also explored the impact on the inferred \mbh\ of assuming the other two MGE descriptions.

\subsection{ALMA Data\label{alma}}

Under Program 2016.1.01010.S, we obtained Cycle 4 ALMA data in Band 6 on 2017 August 19 in the C40$-$6 configuration, with minimum and maximum baselines of 21.0 and 3700 m, respectively.
Observations consisted of a single pointing with one spectral window centered on 225.579 GHz, corresponding to the redshifted frequency of the 230.538 GHz $^{12}$CO(2$-$1) transition, and two spectral windows centered on continuum bands with average observed frequencies of 227.754 GHz and 241.558 GHz.
While we detected dust continuum in the latter spectral windows, here we focus on the CO spectral window.
The total on-source integration time was 34.4 minutes.

We processed the observations using Common Astronomy Software Applications (CASA) version 4.7.2.
The emission-free channels were imaged to produce a continuum map and were used for $uv$-plane continuum subtraction.
A \texttt{TCLEAN} deconvolution with Briggs weighting ($r = 0.5$; \citealt{Briggs1995}) produced a synthesized beam with a full width at half maximum (FWHM) of 0\farcs{197} along the major axis and 0\farcs{104} along the minor axis, with a geometric mean resolution of $0\farcs{143}$, or 63.1 pc in \gal.
The beam PA is $9.271^{\circ}$ east of north.
The data were flux-calibrated using the ALMA standard quasar J$0238+1636$, and we adopt a standard $10\%$ uncertainty in absolute flux calibration at this frequency \citep{Fomalont2014}.

The final data cube has a scale of 0\farcs{02} pixel$^{-1}$ and 120 frequency channels each with a 15.62 MHz width, corresponding to $\sim$20.8 \kms\ relative to the frequency of the CO($2-1$) line at the galaxy's systemic velocity.
The rms noise level of emission-free regions in the cube is $0.3$ mJy beam$^{-1}$ channel$^{-1}$, and we detected CO emission in channels 30 through 78, corresponding to recessional velocities of $5949.0 - 6966.0$ \kms.

\subsubsection{CO(2$-$1) Emission Properties\label{emission}}

\begin{figure*}
\includegraphics[width=\textwidth]{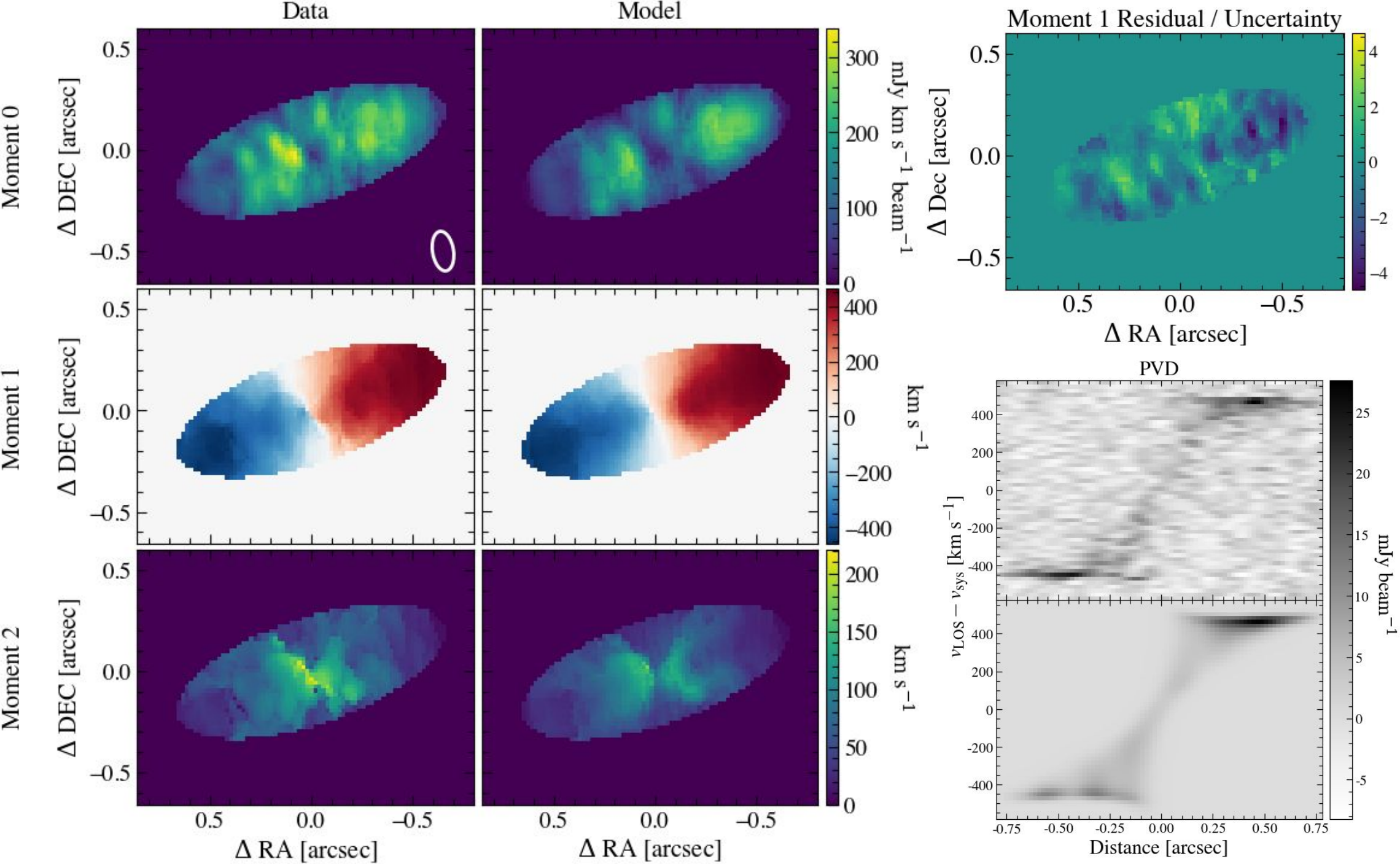}
\caption{Maps of the zeroth, first, and second moments for \gal\ constructed from the ALMA data cube (left) and from the best-fit model cube (middle), shown within our fiducial elliptical fitting radius of 0\farcs{7}.
The maps reveal the disk is in regular rotation.
The residual (data-model) normalized by the uncertainty of the first moment map (top, right) shows the model is a good description of the data, with no significant structure.
The PVD constructed from the data cube (middle, right) and best-fit model cube (bottom, right) are shown with the line-of-sight velocity given relative to the systemic velocity listed in Table \ref{tab_fiducial}.
The PVD was extracted along the disk major axis at a PA of $108.6^\circ$, measured as the degrees east of north to the approaching side of the disk, with an extraction width of 0\farcs{143}, corresponding to the geometric mean width of the synthesized ALMA beam.
The moments, residual moment map, and PVD values are linearly mapped to colors given by the scale bar to the right of each map, and the white ellipse in the top left panel shows the ALMA synthesized beam.
Note that we fit models directly to the data cube, and the moment maps and PVDs represent quantities extracted from the data and best-fit model cubes.}
\label{fig_fiducial_moments}
\end{figure*}

In the left column of Figure \ref{fig_fiducial_moments}, we present spatially resolved maps of the zeroth, first, and second moments of the ALMA data, which correspond to the integrated CO(2$-$1) emission, the projected line-of-sight velocity ($v_{\mathrm{los}}$), and the projected line-of-sight velocity dispersion ($\sigma_{\mathrm{los}}$), respectively.
When constructing these maps, we masked pixels that do not contain CO emission.
As can be seen in Figure \ref{fig_fiducial_moments}, the CO emission traces a regular disk that is $\sim$1\farcs5 in diameter and is co-spatial with the dust (Figure \ref{fig_hst}).
There is a clear dearth of emission within a projected radius of $\sim$0\farcs1 of the nucleus.
The center is not completely devoid of emission, but has $\sim$4$\times$ lower surface brightness than the peak CO emission in the disk.
Such a dearth of emission at the nucleus may be relatively common for molecular gas in ETGs \citep{Boizelle2017,Davis2018,Boizelle2019}.
The velocity map in Figure \ref{fig_fiducial_moments} shows that the southeast side of the disk is blueshifted and the northwest portion is redshifted, with $v_\mathrm{los}$ reaching $\sim$$\pm$460 km s$^{-1}$.
The velocity dispersion peaks at 220 km s$^{-1}$ and the map exhibits the characteristic ``X'' shape, which arises from the combination of rotational broadening and beam smearing in regions with steep velocity gradients.

To further characterize the gas disk kinematics, we fit a harmonic expansion to the $v_\mathrm{los}$ map using the {\tt kinemetry} routine \citep{Krajnovic2006}.
When using kinemetry, we binned adjacent spectra together using Voronoi tessellation \citep{Cappellari2003} to increase the signal-to-noise ratio (S/N) of the $v_\mathrm{los}$ map.
We find that the kinematic PA varies by only $\sim$3$^\circ$ and the axis ratio of the kinematic ellipse changes from $\sim0.6-0.8$ across the disk, indicating that there is no major kinematic twist or significant warp in the disk.
The ratio between the harmonic coefficients $k_5$ and $k_1$ is small ($\lesssim$0.04) without distinct features over the disk's extent, suggesting that the kinematics are dominated by coherent rotation.

Additionally, we construct a position-velocity diagram (PVD) by extracting flux along the semi-major axis of the CO disk at an angle of $108.6^\circ$, measured counterclockwise from north to the blueshifted side of the disk.
This angle matches the gas disk PA found from early dynamical model runs, and is also consistent with the PA found from {\tt kinemetry}.
The extraction width was set to the geometric mean of the ALMA beam FWHMs (0\farcs{143}).

The observed PVD is shown in the middle right-hand panel of Figure \ref{fig_fiducial_moments}.
The PVD displays a continuous distribution of emission across the full range of velocities present in the disk.
The decrease in CO emission at the nucleus seen in the zeroth moment map in Figure \ref{fig_fiducial_moments} is also reflected in the PVD, as the emission at central radii has low S/N.
However, high-velocity emission is persistent within $0\farcs1$ indicating a resolved BH SOI.
The $v_{\mathrm{los}}$ peaks at 480 km s$^{-1}$ at a radius of $0\farcs{08}$ (35 pc).
Assuming the disk is inclined by \inc\ $\approx 68^\circ$, the circular velocity at this radius is $v_c = 517$ km s$^{-1}$, corresponding to an enclosed mass of $\sim 2 \times 10^9\ M_\odot$.
Assuming the enclosed mass is dominated by \mbh, this BH mass corresponds to $r_{\mathrm{SOI}} \sim 0\farcs{2}$ and is in approximate agreement with the expected \mbh\ based on the stellar velocity dispersion reported in \citet{Yildirim2017} and the \msig\ relation \citep{Kormendy2013,Saglia2016}.

The total CO(2$-$1) flux is $4.91 \pm 0.05$ Jy km s$^{-1}$ with an additional 10\% systematic uncertainty in the flux scale.
From the observed flux, we estimate the CO(2$-$1) luminosity ($L^\prime_{\mathrm{CO(2-1)}}$) following \citet{Carilli2013} and convert to a CO(1$-$0) luminosity ($L^\prime_{\mathrm{CO(1-0)}}$) using $R_{21} \equiv L^\prime_{\text{CO}(2-1)}/L^\prime_{\text{CO}(1-0)} = 0.7$ (for an excitation temperature $\sim5-10$ K; e.g., \citealt{Lavezzi1999}).
We determine the H$_2$ mass by assuming a CO-to-H$_2$ conversion factor of $\alpha_{\text{CO}} = 3.1\ M_\odot$ pc$^{-2}$ (K km s$^{-1}$)$^{-1}$ \citep{Sandstrom2013}.
After estimating the H$_2$ gas mass as $M_{\text{H}_2} = \alpha_{\text{CO}} L^\prime_{\mathrm{CO(1-0)}}$, we derived the total gas mass by adopting a helium mass fraction of $f_{\text{HE}} = 0.36$, such that the total gas mass is $M_{\text{gas}} = M_{\text{H}_2} (1+f_{\text{HE}})$.
We find a total gas mass of $(1.47\pm0.01)\times10^8\ M_\odot$ with an additional systematic uncertainty of $\pm0.15\times10^8\ M_\odot$ from the flux calibration.
There are potentially larger systematic uncertainties associated with $\alpha_{\text{CO}}$ (e.g., \citealt{Barth2016b}), as the $\alpha_{\text{CO}}$ we use was calibrated in spiral galaxy disks and may not apply directly to ETG disks, and we consider the gas mass to be a rough estimate.
The gas mass for UGC 2698 has the same order of magnitude as masses found in other ETGs with CO detections \citep{Boizelle2017,Ruffa2019}.

\section{\label{model}Dynamical Modeling}

The ALMA observations reveal a well-ordered circumnuclear CO(2$-$1) disk, and we fit gas-dynamical models to the data assuming the gas participates in circular rotation.
We closely followed the approach described in \citet{Barth2016b} and \citet{Boizelle2019}, which was based on earlier work modeling ionized gas disks observed with HST (e.g., \citealt{Macchetto1997, Barth2001, Walsh2010, Walsh2013}).

Models were constructed on a grid that is oversampled relative to the ALMA spatial pixel by a factor of $s = $ \overs, such that each ALMA 0\farcs02 pixel is divided into $s \times s = 16$ subpixels.
Using the major-axis orientation angle (\pa; defined as degrees east of north to the approaching side of the disk) and the disk inclination angle (\inc), we determined the radius to each pixel relative to the BH location ($x_0$, $y_0$).
We calculated the circular velocity at each radius due to the enclosed mass, which consists of contributions from \mbh\ and the extended stellar mass distribution.
In order to determine the host galaxy's contribution to the circular velocity ($v_{c,\star}$), we deprojected the dust-masked MGE from \S\ref{mge}, assuming axisymmetry and an inclination angle that matches the gas disk \inc, and multiplied by a stellar mass-to-light ratio.
We assumed that the gas disk mass is negligible, but we test this assumption in Section \ref{error}.
We also ignored the contribution from dark matter, which is expected to be insignificant on the scale of the gas disk.
From $v_c$, we determined $v_\mathrm{los}$ using \inc\ and the disk azimuthal angle.

At the subsampled scale, we employed Gaussians to describe the intrinsic model line profiles.
We centered the line profiles on $v_\mathrm{los}$ at each point on the model grid and set the widths of the line profiles to be equal to an intrinsic turbulent velocity dispersion ($\sigma_\mathrm{turb}$).
This velocity dispersion is the additional line width needed to match the observations once accounting for beam smearing and unresolved rotation, and it is not assigned a dynamical origin.
We examined two parameterizations for $\sigma_\mathrm{turb}$.
For the fiducial model, $\sigma_\mathrm{turb}$ is a constant (\sig) throughout the disk, but in Section \ref{error} we further explore a $\sigma_\mathrm{turb}$ profile that varies with radius.
Both the line centers and widths were converted to frequency using the systematic velocity ($v_\mathrm{sys}$).
We constructed the Gaussian line profiles on the same frequency axis as the observed ALMA cube, with a channel spacing of 15.62 MHz.

We weighted the model line profiles with an estimate of the intrinsic CO distribution.
To approximate the intrinsic CO surface brightness, we collapsed the observed ALMA data cube by integrating over the frequency axis while masking noise in each slice.
The resulting image was deconvolved with the ALMA beam, using 10 iterations of the \texttt{lucy} task \citep{Richardson1972,Lucy1974} in the \texttt{scikit-image} package \citep{Walt2014}.
As we do not have information about the intrinsic CO surface brightness on subpixel scales, we assumed the flux in each native pixel is equally divided among the $s \times s$ subpixels.
We also included a scale factor, \fw, when weighting the line profiles to account for any slight normalization mismatch between the data and model.

With the above prescription, we produced an intrinsic model cube on subpixel scales.
We then averaged each set of $s \times s$ subpixels to create a model line profile at each native ALMA pixel.
Thus, asymmetric line profiles can be seen in the rebinned model cube, even though the line profiles are Gaussian on subpixel scales.
Finally, we convolved each frequency slice of the model cube with the synthesized ALMA beam.

Our fiducial model has nine free parameters: the BH mass \mbh, the mass-to-light ratio \ml, the inclination \inc, the disk position angle \pa, the systemic velocity \vsys, the intrinsic turbulent velocity dispersion \sig, the BH location (\xloc, \yloc), and the line profile scale factor \fw, and we optimized them using the nested sampling code {\tt dynesty} \citep{Speagle2019}.
We directly compared the model cube to the observed data cube after down-sampling both the model and data cubes in bins of $\ds\times\ds$ spatial pixels to mitigate correlated noise, following \citet{Barth2016b}.
We adopted a likelihood of $L \propto \exp(-\chi^2/2)$ with $\chi^2 = \sum_{i} ((d_i - m_i)^2/\sigma_i^2)$, where $d_i$ and $m_i$ are the down-sampled data and down-sampled model cube values, respectively, and $\sigma_i$ is the noise, which is assumed to be the same for all pixels within a single velocity channel.
We estimated the noise in each velocity channel by calculating the standard deviation in an emission-free region of the down-sampled data cube near the CO disk.
We calculated $\chi^2$ within a fixed region that spans 57 velocity channels (corresponding to $|v_\mathrm{los}-v_\mathrm{sys}|\lesssim 590$ km s$^{-1}$) and over an ellipse with a semi-major axis of $r_{\text{fit}} = 0\farcs{70}$, an axis ratio $q_{\text{ell}} = 0.38$, and a position angle \pa$_{\text{ell}} = 109^\circ$ east of north.
This fitting region encompasses nearly all of the CO emission without including excess noise.
There are 5301 data points and 5292 degrees of freedom.
We also tested using different $r_{\text{fit}}$ values, which will be discussed in Section \ref{error}. 

When running {\tt dynesty}, we used 250 live points and a threshold of 0.02.
Live points are a nested sampling parameter that sample the prior, and the point with the lowest likelihood is iteratively replaced with a new point at an improved likelihood.
The threshold determines when to stop sampling and initially corresponds to the log-ratio between the current estimated Bayesian evidence and the evidence that remains to be sampled.
After the initial sampling stage reaches the threshold, batches of nested samples are added until the fractional error on the posterior reaches another threshold, which we also set to 0.02.
We tested a wide range of numbers of live points and initial thresholds and confirmed that our results are unchanged.
We employed flat priors, such that all free parameters were sampled uniformly in linear space, except for \mbh, which was sampled uniformly in logarithmic space.
Our prior ranges were chosen to be much larger than the expected posterior widths.
We took the $1\sigma$ and $3\sigma$ uncertainties to be the $68\%$ and $99.7\%$ confidence intervals of the parameter posterior distributions, respectively.

\section{\label{results}Modeling Results}

When fitting the rotating thin-disk models to the ALMA data cube, we find that \mbh\ $ = (2.46\pm0.07[^{+0.21}_{-0.19}]) \times 10^9\ M_\odot$ and \ml\ $ = 1.70\pm0.01[\pm0.04]$ $M_\odot/L_\odot$ ($1\sigma$ and [$3\sigma$] uncertainties), with $\chi^2 = 6496.0$ and a reduced $\chi^2$ ($\chi^2_\nu$) of $1.228$.
\citet{Yildirim2017} fit orbit-based dynamical models to large-scale stellar kinematics and found \ml\ $ = 1.80\pm0.10$ $M_\odot/L_\odot$ for \gal, which is consistent with our \ml.
Our \ml\ is also similar to predictions from simple stellar population models \citep{Vazdekis2010}, which suggest \ml\ $ = 1.50-1.80$ $M_\odot/L_\odot$ for a \citet{Salpeter1955} initial mass function (IMF) and \ml\ $ = 1.10-1.20$ $M_\odot/L_\odot$ for a \citet{Kroupa2001} IMF with solar metallicity and a $\sim10-12$ Gyr stellar age.

\begin{figure}
\includegraphics[width=0.47\textwidth]{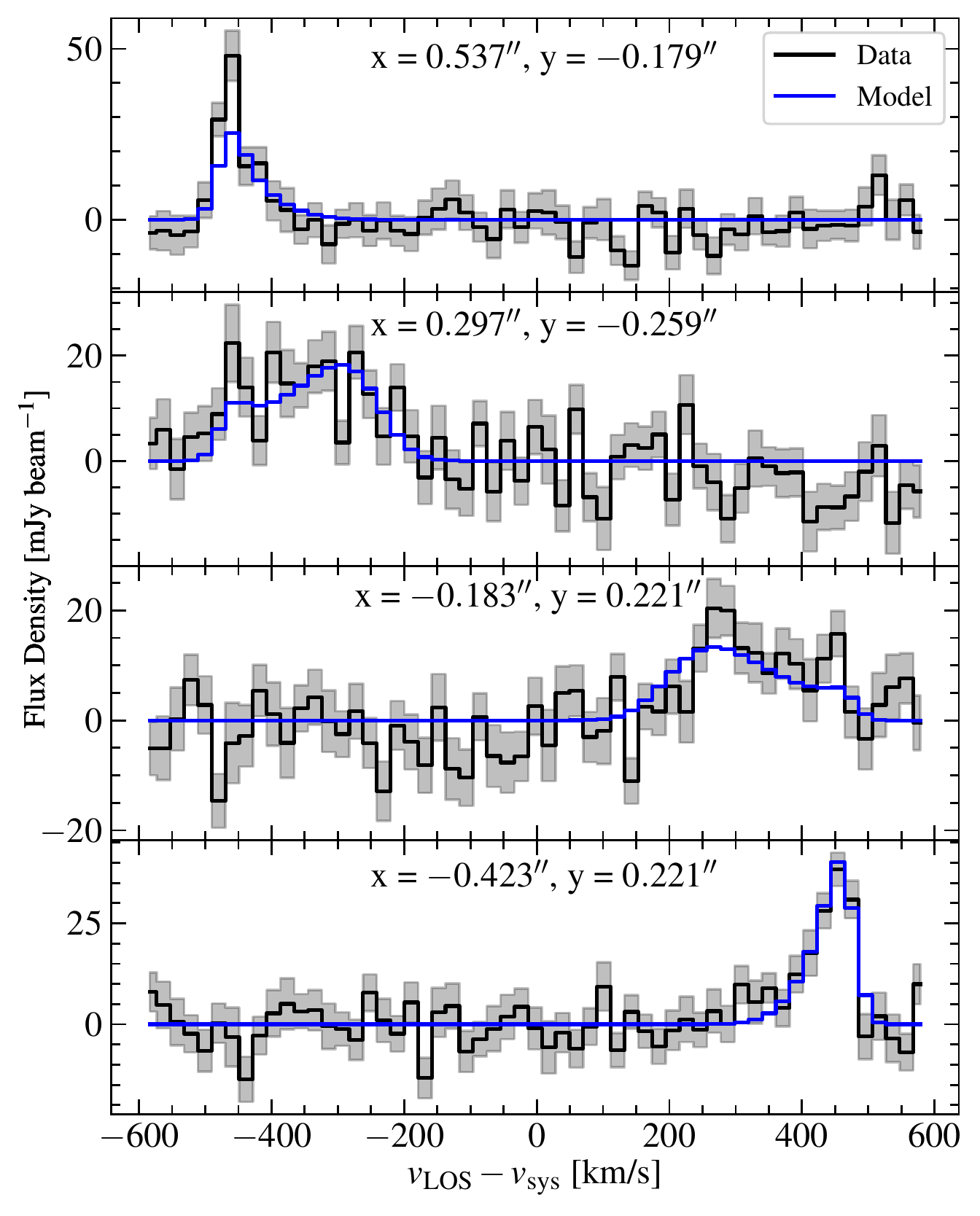}
\caption{Line profiles from the down-sampled data and model cubes at four different disk locations, with the ($x,y$) position given relative to (\xloc, \yloc), where +y corresponds to north and +x corresponds to east.
The shaded gray bands show the noise in each velocity channel.
On the blueshifted side of the disk (e.g., in the top panel), the model tends to underestimate the observed flux; however, the model matches the observed velocity quite well.}
\label{fig_fiducial_lps}
\end{figure}

The moment maps and PVD constructed from the best-fit model are presented in Figure \ref{fig_fiducial_moments}.
In addition, we provide a residual map comparing the observed first moment and model, divided by the first moment uncertainty.
We determined the first moment uncertainty using a Monte Carlo simulation, in which we generated 1000 mock observed data cubes with pixel values randomly drawn from a Gaussian distribution centered on the fiducial model cube and with a width equal to the the standard deviation of an emission-free region of the given velocity channel in the observed data cube.
After constructing the first moment map during each of the 1000 realizations, we took the standard deviation of the resultant maps as the uncertainty of the first moment.
We also show example line profiles and the best-fit model in Figure \ref{fig_fiducial_lps}.

\begin{figure*}
\includegraphics[width=\textwidth]{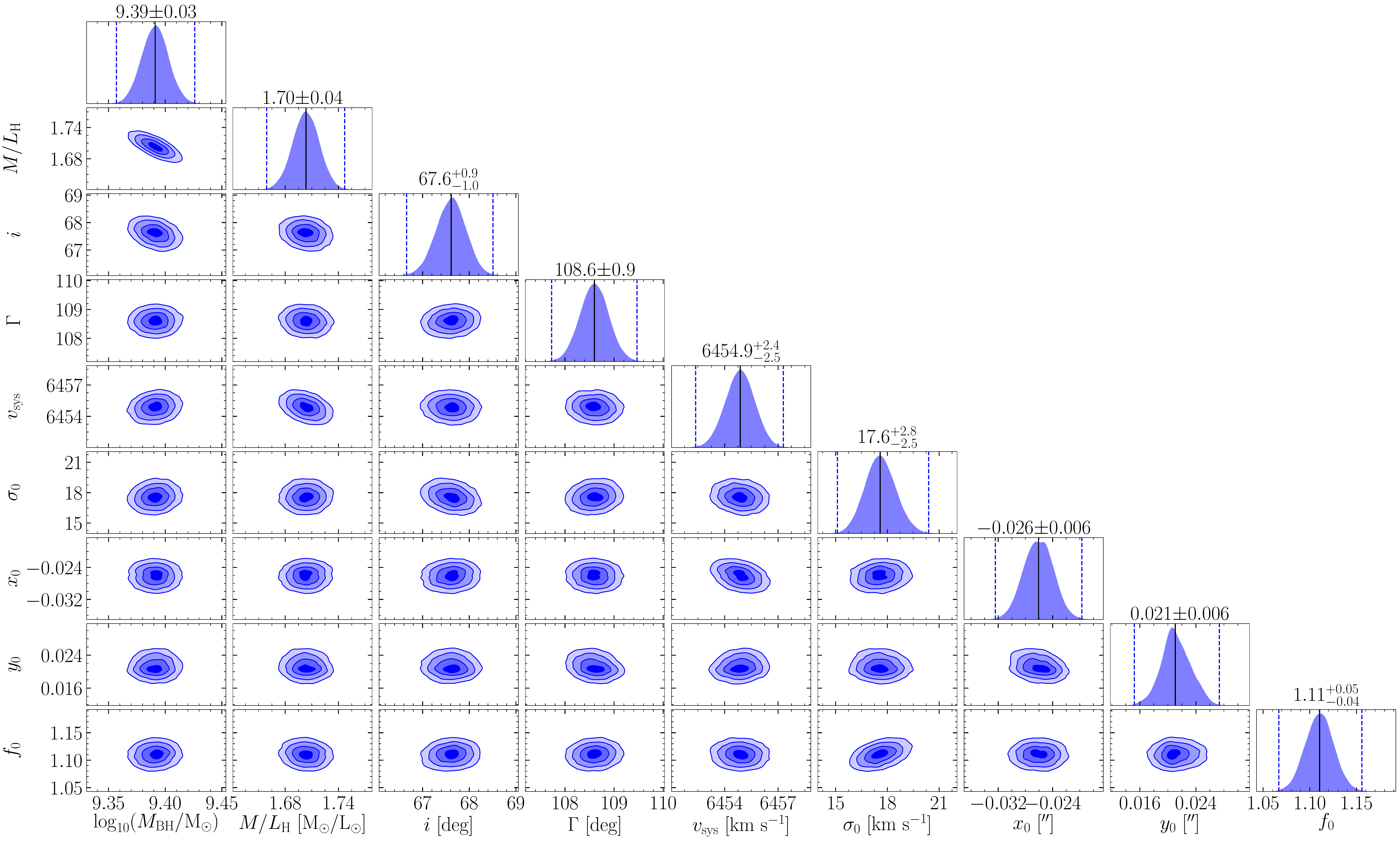}
\caption{Corner plot displaying the one-dimensional (1D; top panels) and 2D posteriors of the free parameters in our model.
For each parameter, the median is shown as a black solid line and the $3\sigma$ confidence intervals are shown as dashed blue lines in the 1D posterior plots.
The median and $3\sigma$ uncertainties for each parameter are listed above each 1D posterior.
The contours in the 2D posterior panels correspond to $0.5\sigma, 1\sigma, 1.5\sigma$, and $2\sigma$ confidence levels.
We find a BH mass of \mbh\ $= (2.46\pm0.07^{[+0.21]}_{[-0.19]}) \times 10^9$ $M_\odot$, listed here with $1\sigma$ ($3\sigma$) statistical uncertainties.
Generating a model using the medians of each posterior and comparing to the data results in $\chi^2_\nu = 1.228$.
The ALMA observations marginally resolve the BH SOI, and there is a clear correlation between \mbh\ and \ml.
As a result, the manner in which we treat the stellar light profile is the largest source of systematic uncertainty in \mbh.}
\label{fig_fiducial_corner}
\end{figure*}

Figures \ref{fig_fiducial_moments} and \ref{fig_fiducial_lps} show that the model is a good fit to the data, although the model modestly underestimates the observed flux on the approaching side of the disk, $\sim$0\farcs{1} from the center.
Previous work has found that the choice of the intrinsic flux map can impact the quality of the fit, but does not have a significant effect on the inferred \mbh\ \citep{Marconi2006,Walsh2013,Boizelle2021}.
All of the best-fit parameters are given in Table \ref{tab_fiducial}, and the posterior distributions are shown in Figure \ref{fig_fiducial_corner}.

\begin{deluxetable}{ccccc}[t]
\tabletypesize{\small}
\tablecaption{UGC 2698 Model Results}
\tablewidth{0pt}
\tablehead{
\colhead{Parameter} & 
\colhead{Median} & 
\colhead{$1\sigma$} &
\colhead{$3\sigma$} &
\colhead{Prior range} \\[-1ex]
\colhead{(1)} & 
\colhead{(2)} & 
\colhead{(3)} &
\colhead{(4)} &
\colhead{(5)}
}
\startdata
\mbh\ [$10^9\ M_\odot$] & ${2.46}$ & $\pm0.07$ & $_{-0.19}^{+0.21}$ & ${0.10} \rightarrow {10.00}$ \\
\ml\ [$M_\odot/L_\odot$] & ${1.70}$ & $\pm0.01$ & $\pm0.04$ & ${0.30} \rightarrow {3.00}$ \\
\inc\ [$^\circ$] & ${67.6}$ & $\pm0.3$ & $_{-1.0}^{+0.9}$ & ${52.4} \rightarrow {89.0}$ \\
\pa\ [$^\circ$] & ${108.6}$ & $\pm0.3$ & $\pm0.9$ & ${95.0} \rightarrow {125.0}$ \\
\vsys\ [km s$^{-1}$] & ${6454.9}$ & $\pm0.8$ & $_{-2.5}^{+2.4}$ & ${6405.0} \rightarrow {6505.0}$ \\
\sig\ [km s$^{-1}$] & ${17.6}$ & $\pm0.9$ & $_{-2.5}^{+2.8}$ & ${0.0} \rightarrow {40.0}$ \\
\xloc\ [\arcsec] & ${-0.026}$ & $\pm0.002$ & $\pm0.006$ & ${-0.031} \rightarrow {0.049}$ \\
\yloc\ [\arcsec] & ${0.021}$ & $\pm0.002$ & $\pm0.006$ & ${-0.038} \rightarrow {0.042}$ \\
\fw\ & ${1.11}$ & $_{-0.02}^{+0.01}$ & $_{-0.04}^{+0.05}$ & ${0.50} \rightarrow {1.50}$ \\
\enddata
\tablecomments{Results from the best-fit gas-dynamical model.
Column (1) lists the model parameters, column (2) shows the median of the posterior distribution, column (3) lists the $1\sigma$ statistical uncertainties, column (4) shows the $3\sigma$ statistical uncertainties, and column (5) provides the prior range for the nested sampling.
The dust-masked MGE cannot be deprojected for inclination angles below $52.4^\circ$, so we took \inc $=52.4^\circ$ as the lower bound on the prior.
The \xloc\ and \yloc\ parameters are measured in arcseconds relative to the centroid of the continuum emission, which is at RA = $3^{\mathrm{h}} 22^{\mathrm{m}} 2.8896^{\mathrm{s}}$ and Dec = $+40^\circ 51\arcmin 50\farcs{0382}$ (J2000).
Positive \xloc\ and \yloc\ values correspond to shifts to the east and north, respectively.}
\label{tab_fiducial}
\end{deluxetable}

\subsection{Error Budget\label{error}}

While the statistical uncertainties from the fits with {\tt dynesty} are small (about a few percent of \mbh), there are several other possible sources of uncertainty that are tied to choices we made when constructing our model.
Therefore, we performed a number of tests in order to assess the impact on \mbh, which we describe below.

\textit{Dust extinction.}
We optimized the fiducial gas-dynamical model using the dust-masked MGE from Section \ref{mge}.
When instead using the original MGE, which ignores the presence of dust, we found \mbh\ $ = 1.71\times10^9\ M_\odot$, representing a $\sim$31\% decrease from the fiducial \mbh, and an increased \ml\ of 1.94 $M_\odot/L_\odot$.
Conversely, when using the dust-corrected MGE to determine $v_{c,\star}$, we found that the BH mass increased to \mbh\ $ = 3.15\times10^9\ M_\odot$, corresponding to a $\sim$28\% change from the fiducial value, and a smaller \ml\ of 1.58 $M_\odot/L_\odot$.
Both models are worse fits to the data compared to the fiducial model's $\chi^2 = 6496.0$ and $\chi^2_\nu = 1.228$.
The dynamical model with the original MGE returns $\chi^2 = 6551.3$ and $\chi^2_\nu = 1.238$, while the model using the dust-corrected MGE yields $\chi^2 = 6518.4$ and $\chi^2_\nu = 1.232$.

We selected the dust-masked MGE to use in our fiducial model for two reasons.
First, the dust-masked MGE accounts for the fact that there is dust present, while the original MGE ignores dust and the dust-corrected MGE introduces additional assumptions and uncertainties during the correction process.
Second, of the three MGE models, the dust-masked MGE yields the best-fitting model.

\textit{Gas mass.}
Our primary dynamical model neglects the mass of the molecular gas disk.
However, we performed a run that included the contribution of the gas disk itself to the gravitational potential.
From the zeroth moment map, we measured the CO surface brightness as a function of radius within elliptical annuli, determined the associated projected surface mass densities, and integrated assuming a thin disk \citep{Binney2008} to determine the contribution to the circular velocity due to the gas ($v_{c,\mathrm{gas}}$).
We found a maximum $v_{c,\mathrm{gas}}$ of $\sim$42 km s$^{-1}$, which is only $\sim$8\% of the maximum circular velocity due to the stars ($v_{c,\star} = 512$ km s$^{-1}$).
When incorporating $v_{c,\mathrm{gas}}$ alongside our model $v_{c,\star}$, we find that the best-fit parameters and uncertainties are essentially identical to those in our fiducial model, and the total $\chi^2$ increases by 2.3 from the fiducial model.
Therefore, the inclusion of the gas mass does not affect our results.

\textit{Turbulent velocity dispersion.}
In addition to adopting a constant turbulent velocity dispersion, we tested an exponential profile with three free parameters, defined as $\sigma_{\text{turb}} = \sigma_0 + \sigma_1 \exp(-r / r_0)$.
When this model was optimized, $\sigma_0$ and $\sigma_1$ were both poorly constrained.
The $\sigma_0$ posterior probability distribution peaked at $\sim$5 km s$^{-1}$, but with $3\sigma$ uncertainties spanning the full prior range from $0-100$ km s$^{-1}$.
The $\sigma_1$ parameter was marginally better constrained, with the distribution peaking more strongly at $\sim$10 km s$^{-1}$, although the $3\sigma$ uncertainties were still very large ($0 - 92$ km s$^{-1}$).
On the other hand, $r_0$ was well constrained, with $r_0 = 17.54^{+2.78}_{-2.39}$ pc ($3\sigma$ uncertainties).
This $r_0$ is equivalent to $\sim$2 ALMA pixels, or $\sim$0\farcs{04}.
The best-fit \mbh\ is $0.3\%$ larger than the fiducial model \mbh.
All other best-fit parameters and uncertainties were unchanged from those in the fiducial model.
The S/N of the ALMA data is too low to warrant exploration of other spatially varying turbulent velocity dispersion profiles.

\textit{Radial motion.}
Although we found no evidence for non-circular motion in the \gal\ disk (e.g., there is no significant kinematic twist in the first moment map), we introduced two toy models to determine how much radial motion is allowed by the ALMA data and the possible impact on \mbh.
For the first model, we followed \cite{Boizelle2019} and included a simple radial velocity term ($v_{\text{rad}}$) that is constant with radius.
This component is projected into the line-of-sight, added to $v_{\text{LOS}}$, and optimized alongside the other free parameters in the model.
We found a small outflow of $14.3^{+10.9}_{-11.4}$ km s$^{-1}$ ($3\sigma$ uncertainties), with all other free parameters consistent within $1\sigma$ with those found for the fiducial model.
The best-fit \mbh\ decreases by 1.2\% from the fiducial value, and $\chi^2_\nu$ modestly increases to 1.233.

For the second model, we allowed the radial velocity to vary with radius, following \citet{Jeter2019}, using a free parameter $\kappa$.
The $\kappa$ parameter is multiplied by $v_c$, projected along the line-of-sight, and added to $v_{\text{LOS}}$.
We allowed for the possibility of inflow or outflow by letting $\kappa$ vary between -1 and 1.
When optimized, the model shows a preference for a slight outflow with $\kappa = 0.03^{+0.02}_{-0.02}$ ($3\sigma$ uncertainties).
With $\kappa = 0.03$, $99.7\%$ of the radial velocities over the gas disk are between $0.1 - 14.4$ km s$^{-1}$, with a maximum radial velocity of $59.0$ km s$^{-1}$ at a radius of $1.61$ pc.
The median radial velocity across the entire disk is $12.4$ km s$^{-1}$, which is similar to the constant $v_{\text{rad}}$ model result.
Despite the adjustments to $v_{\text{LOS}}$, all parameter values, including \mbh, are consistent with those from the fiducial model within $1\sigma$.
The best-fit \mbh\ decreases by 1.1\% from the fiducial value, and $\chi^2_\nu$ modestly increases to 1.232.

\textit{Oversampling factor.}
For the fiducial model, we used a spatial pixel oversampling factor of $s = \overs$.
Previous ionized gas-dynamical (e.g., \citealt{Barth2001}) and molecular gas-dynamical models (e.g., \citealt{Boizelle2019}) have found that \mbh\ can depend on the oversampling factor, varying by up to a few percent.
Therefore, we tested using $s =$ 1, 2, 3, 6, 8, 10, and 12.
We found that the best-fit \mbh\ and formal model fitting uncertainties change very little.
The model with $s = 1$ produces the largest change in \mbh, with \mbh$ = 2.51\times10^9\ M_\odot$ ($1.8\%$ larger than the fiducial model) and $\chi^2_\nu = 1.225$, which is a slight improvement over the fiducial model $\chi^2_\nu$ of 1.228.
The models with $s \geq 3$ result in \mbh\ shifts of $\leq$0.03\% from the fiducial \mbh.

\textit{Fitting ellipse.}
The size of the fitting region determines which down-sampled data and model pixels are compared.
While our fiducial model uses an ellipse with a semi-major axis of 0\farcs{7}, we also examined models with $r_{\text{fit}} =$ 0\farcs{5}, 0\farcs{6}, and 0\farcs{8}.
We did not test models with $r_{\text{fit}} > $ 0\farcs{8} because that would incorporate numerous pixels outside of the emission-line disk in our calculation of $\chi^2$.
For $r_{\text{fit}} < $ 0\farcs{5}, there are not enough down-sampled pixels to constrain the model.
The best-fit \mbh\ increases with fitting radius, from \mbh\ $=2.27\times10^9\ M_\odot$ for $r_{\text{fit}} = 0\farcs{5}$ (corresponding to a $-7.8$\% change from the fiducial value) to \mbh\ $ = 2.53\times10^9\ M_\odot$ for $r_{\text{fit}} = 0\farcs{8}$ (a $2.9$\% shift).
Nevertheless, the masses remain consistent within $3\sigma$ of the fiducial \mbh\ value.
Likewise, the other model parameters show slight shifts, but are all well within the $3\sigma$ statistical uncertainties of the fiducial model.

\textit{Down-sampling factor.}
We used $\ds\times\ds$ spatial pixel down-sampling to compare the data and model cubes.
However, the ALMA beam is elongated with a FWHM of 0\farcs{104} (5.2 pixels) along the minor axis and a FWHM of 0\farcs{197} (9.9 pixels) along the major axis, which is nearly aligned with the $y$-axis of the UGC 2698 cube.
Therefore, we tested down-sampling factors that were $2\times$ larger in $y$ than in $x$, including adopting bins of $4\times8$ and $5\times10$ spatial pixels.
In both cases, the best-fit \mbh\ and \ml\ were consistent with the fiducial model results within the $1\sigma$ uncertainties.
When we used $4\times8$ pixel ($5\times10$ pixel) down-sampling, \mbh\ decreased by $1.2\%$ ($1.5\%$) from the fiducial value.

\textit{Optimization parameters.}
We modified the parameters used with {\tt dynesty} and confirmed that our results remained unchanged.
We tested increasing the number of live points used in the nested sampling code from 250 points to 1000 points.
This run yielded best-fit parameter values identical to those of the fiducial model.
In order to ensure that our model was sufficiently converged, we explored decreasing the sampling threshold from 0.02 to 0.001.
All parameters were again the same as in the fiducial model, with a shift in \mbh\ of $<$0.1$\%$.
We also ran {\tt dynesty} using much wider priors, allowing for the full range of physically possible values for each free parameter.
This model had the same best-fit parameters as the fiducial model, again with a $<$0.1$\%$ change in \mbh.

\textit{Intrinsic flux map.}
Varying the number of iterations in the Lucy-Richardson deconvolution process also produced essentially no change in our results.
We tested using five and 15 rather than the fiducial ten Lucy-Richardson deconvolution iterations.
When incorporating these two alternative intrinsic flux maps into the dynamical models, we found best-fit parameters that agreed with the fiducial model within the $1\sigma$ uncertainties.
Using the intrinsic flux maps produced from five (15) deconvolution iterations resulted in a best-fit \mbh\ that decreased by $1.3\%$ (increased by $0.6\%$) relative to the fiducial model.

\textit{Final error budget}.
Adding the \mbh\ changes described above in quadrature, we find a systematic (sys) uncertainty at the 30\% level, which is almost entirely due to the treatment of dust when constructing the MGE.
This systematic uncertainty is larger than the 3$\sigma$ statistical (stat) uncertainty derived from the posterior probability distribution.
To summarize, the UGC 2698 \mbh\ is $(2.46\pm0.07$ [stat, $1\sigma$] $^{+0.21}_{-0.19}$ [stat, $3\sigma$] $^{+0.70}_{-0.78}$ [sys]) $\times 10^9\ M_\odot$.

\section{\label{discussion}Discussion}

Our ALMA-based measurement is the first dynamical \mbh\ determination for UGC 2698, and it is the only precision gas-dynamical constraint for any of the local compact galaxies from \citet{Yildirim2017}.
Although \citet{Scharwachter2016} examined CO(1$-$0) emission in the nearby compact galaxy NGC 1277, with 1\arcsec\ and 2\farcs9-resolution IRAM Plateau de Bure Interferometer observations they were unable to distinguish between a $\sim5\times10^9\ M_\odot$ BH and a $\sim2\times10^{10}\ M_\odot$ BH.
ALMA, with its increased resolution and sensitivity, provides an opportunity to examine the detailed kinematic structure of molecular gas disks on much smaller scales and acquire robust gas-dynamical BH masses for the local compact galaxies.
Below, we examine the UGC 2698 BH SOI (Section \ref{resolution}), we compare \gal\ to other local compact galaxies and to the BH$-$host galaxy relations (Section \ref{compare_bh}), and we discuss the implications of our measurement for BH$-$galaxy co-evolution (Section \ref{scaling_relations}).

\subsection{\label{resolution}The BH Sphere of Influence}

In order to estimate the BH SOI for UGC 2698, we find the radius where \mbh\ is equal to the enclosed stellar mass.
For our fiducial model from Section \ref{results}, this occurs at $r_{\text{SOI}} = 0\farcs{17}$, which is equivalent to 75 pc.
If we instead use the original MGE and the dust-corrected MGE, and their corresponding inferred BH masses, the radii are 0\farcs{12} (53 pc) and 0\farcs{23} (102 pc), respectively.
For comparison, we also calculated the BH SOI using $r_{\text{SOI}} = G$\mbh$/\sigma_\star^2$, with $\sigma_\star = 304$ km s$^{-1}$ \citep{Yildirim2017}.
Taking the fiducial \mbh\ of $2.46\times10^9\ M_\odot$, we found $r_{\text{SOI}} = 0\farcs{26}$ (114 pc).
With the \mbh\ determined from gas-dynamical models using the original $H$-band MGE and the dust-corrected MGE, $r_{\text{SOI}} = $ 0\farcs{18} (80 pc) and 0\farcs{33} (147 pc), respectively.

Our ALMA observations marginally resolve the BH SOI.
Thus, it is not surprising that we see a clear degeneracy between \mbh\ and \ml\ in Figure \ref{fig_fiducial_corner} and that the systematic errors on \mbh\ are dominated by the adopted stellar light distribution and the treatment of circumnuclear dust.
The ALMA beam is $0\farcs104\times0\farcs197$ (FWHM) with a geometric mean of 0\farcs143.
Following \citet{Rusli2013b}, we compare the BH SOI, as measured using the radius where \mbh\ equals the enclosed stellar mass from the fiducial model, to the average ALMA beam size via $\xi = 2r_{\text{SOI}}/\theta_{\text{FWHM}}$.
We find that $\xi = 2.4$.
\citet{Davis2014} argue that $\xi \sim 2$ is sufficient to make a molecular gas-dynamical BH mass measurement, although our results indicate that it is imperative to account for systematic uncertainties in regimes where the BH SOI is not highly resolved.

With knowledge of the UGC 2698 CO surface brightness on sub-arcsecond scales, follow-up ALMA observations at higher angular resolution can be planned, which would limit the \mbh$-$\ml\ degeneracy and mitigate the dominant source of systematic uncertainty.
As an extreme example of ALMA's capabilities, observations of the massive elliptical NGC 3258 have an extraordinary $\xi\sim17$, allowing gas-dynamical models to pin down \mbh\ to percent-level precision \citep{Boizelle2019}.

\subsection{\label{compare_bh} Comparison to Other Local Compact Galaxies and the BH Scaling Relations}

UGC 2698 is part of a sample of nearby compact galaxies \citep{Yildirim2017}, and three of these galaxies have dynamical \mbh\ measurements in the literature.
NGC 1277, in particular, has been extensively studied (e.g., \citealt{Bosch2012,Emsellem2013,Yildirim2015,Graham2016a,Scharwachter2016}), and high resolution adaptive optics (AO) Gemini/NIFS observations coupled with orbit-based, stellar-dynamical models \citep{Walsh2016} and anisotropic Jeans models \citep{Krajnovic2018} suggest \mbh\ $\sim 5\times10^9\ M_\odot$.
Based on AO Gemini/NIFS data and stellar-dynamical models, NGC 1271 and Mrk 1216 host similarly large BHs, with masses of $\sim (3-5)\times10^9\ M_\odot$ \citep{Walsh2015,Walsh2017}.

With \mbh\ $= 2.46\times10^9\ M_\odot$, UGC 2698 also falls at the upper end of the BH mass distribution.
To place UGC 2698 on the BH scaling relations, we used $\sigma_\star = 304\pm6$ km s$^{-1}$, which is the stellar velocity dispersion within a circular aperture containing half of the galaxy light \citep{Yildirim2017}.
We then measured the total $K$-band luminosity by summing the components of the dust-masked MGE to determine the $H$-band luminosity, assuming an absolute $H$-band ($K$-band) magnitude of 3.37 mag (3.27 mag) for the Sun \citep{Willmer2018} and $H - K = 0.2$ mag for the galaxy \citep{Vazdekis2010}.
We found $L_H = 2.16\times10^{11}\ L_\odot$, corresponding to $L_K = 2.37\times10^{11}\ L_\odot$.
In addition, multiplying the total $H$-band luminosity by the best-fit \ml\ from the fiducial dynamical model ($1.70\ M_\odot/L_\odot$), we found a total stellar mass of $M_{\star} = 3.68\times10^{11}\ M_\odot$.
Using the original MGE, without masking dust, resulted in $L_H = 2.16\times10^{11}\ L_\odot$ ($L_K = 2.37\times10^{11}\ L_\odot$) and $M_\star = 4.19\times10^{11}\ M_\odot$, while the dust-corrected MGE led to $L_H = 2.18\times10^{11}\ L_\odot$ ($L_K = 2.39\times10^{11}\ L_\odot$) and $M_\star = 3.43\times10^{11}\ M_\odot$.

\begin{figure}
\centering
\includegraphics[width=0.47\textwidth]{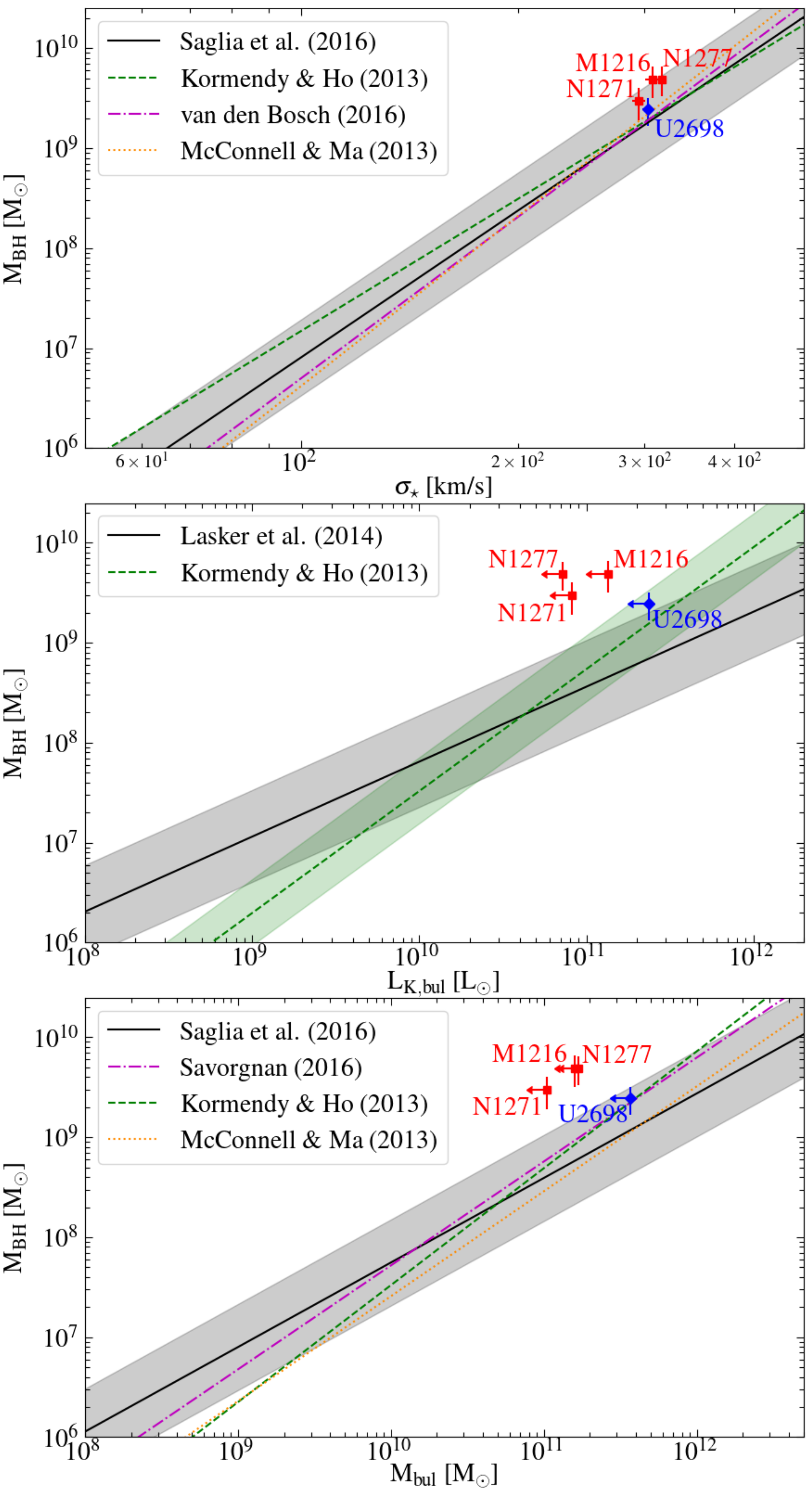}
\caption{\msig\ (top), \mlum\ (middle), and \mmass\ (bottom) scaling relations, including their intrinsic scatter (shaded regions).
Our ALMA-based \mbh\ for UGC 2698 and its systematic uncertainty are shown (blue diamond), along with the AO stellar-dynamical \mbh\ measurements for similar compact galaxies (red squares; \citealt{Walsh2015,Walsh2016,Walsh2017}).
All four galaxies are consistent with \msig.
In the middle (bottom) panel, we conservatively show the total $K$-band luminosity (stellar mass) from the dust-masked MGE for UGC 2698.
Also plotted are the total $K$-band luminosities and stellar masses for NGC 1271, NGC 1277, and Mrk 1216.
Since we use total luminosities and stellar masses rather than bulge measurements, these estimates are upper bounds.
The three previously-measured stellar-dynamical BH masses are positive outliers from the \mlum\ and \mmass\ relations, but \gal\ is consistent with both relations within their intrinsic scatter.}
\label{fig_scaling_relations}
\end{figure}

In Figure \ref{fig_scaling_relations}, we show \gal\ with respect to \msig, \mlum, and \mmass.
Also plotted are NGC 1271, NGC 1277, and Mrk 1216.
The $\sigma_\star$ values come from \citet{Yildirim2017}.
Since UGC 2698 is classified as an elliptical galaxy \citep{Vaucouleurs1991} and there is disagreement in the literature regarding the bulge properties of the local compact galaxy sample \citep{SavorgnanGraham2016,Graham2016b}, we conservatively adopted total galaxy quantities when placing the objects on the BH scaling relations.
The total $K$-band luminosities and stellar masses come from the MGEs for NGC 1271 \citep{Walsh2015}, NGC 1277 \citep{Yildirim2017}, and Mrk 1216 \citep{Yildirim2015}, assuming an absolute $H$-band magnitude of 3.37 mag and $K$-band magnitude of 3.27 mag for the Sun \citep{Willmer2018}, $H-K = 0.2$ mag for old stellar populations \citep{Vazdekis2010}, and the dynamically determined \ml\ values (\citealt{Walsh2015,Walsh2016,Walsh2017}).
We find that UGC 2698 is consistent with the \msig, \mlum, and \mmass\ relations to within their intrinsic scatter.
This result is a departure from the three other compact galaxies, which are consistent with \msig\ but lie above the \mlum\ and \mmass\ relations, even when using total luminosities and stellar masses.
Interestingly, \citet{Zhu2021} re-calculated the BH scaling relations based on classical bulges and the cores of elliptical galaxies, since $z\sim2$ red nuggets are believed to form the cores of local ellipticals, and NGC 1271, NGC 1277, and Mrk 1216 are less significant outliers from their BH mass$-$core mass relation. 

Relative to the three previously-studied compact galaxies and the remaining galaxies in \cite{Yildirim2017}, UGC 2698 is the largest in luminosity and stellar mass.
It has the largest effective radius and exhibits a more extended stellar mass surface density profile.
All of the compact galaxies in the sample are fast rotators \citep{Emsellem2011}, but UGC 2698 is the slowest rotating galaxy and barely meets the threshold to be classified as a fast rotator.
Therefore, UGC 2698 appears to have grown with respect to the rest of the sample and may have experienced dry minor mergers that built up its mass and size, as well as at least one intermediate-to-major dry merger that slowed its rotation \citep{Yildirim2017}.

Comparing the results for NGC 1271, NGC 1277, and Mrk 1216 to that for \gal\ is complicated by the fact that stellar-dynamical and gas-dynamical measurements do not always agree.
In fact, there are only a few galaxies that have both stellar-dynamical and gas-dynamical BH mass measurements, and estimates can differ by factors of $2-3$, with the stellar-dynamical determination usually larger than the gas-dynamical mass \citep{Kormendy2013}. 

Comparisons between stellar-dynamical and molecular gas-dynamical methods, specifically, have found consistent \mbh\ values for NGC 4697 \citep{Schulze2011,Davis2017}, but inconsistent results for NGC 524 and NGC 1332.
The stellar-dynamical \mbh\ for NGC 524 \citep{Krajnovic2009} is $\sim$2$\times$ more massive than the gas-dynamical \mbh\ \citep{Smith2019}, although \citet{Smith2019} note that the BH masses are consistent within $3\sigma$ uncertainties.
NGC 1332 closely resembles the objects in our local compact galaxy sample, displaying a small effective radius for its stellar mass.
The stellar-dynamical \mbh\ for NGC 1332 was found to be over-massive compared to the \mlum\ scaling relation and consistent with \msig\ \citep{Rusli2011}.
However, the ALMA gas-dynamical \mbh\ for NGC 1332 is $\sim$2$\times$ smaller and consistent with both relations \citep{Barth2016a,Barth2016b}.
Furthermore, there is some uncertainty in the galaxy decompositions \citep{Kormendy2013,SavorgnanGraham2016}, leading to conflicting measurements of the bulge luminosity of NGC 1332.
More stellar- and gas-dynamical BH mass comparison studies are needed to better determine the intrinsic scatter in the BH scaling relations, as well as whether galaxies deviating from the relations are significant outliers.

\subsection{\label{scaling_relations} Implications for BH$-$Galaxy Growth}

The local compact galaxies, including UGC 2698, share many properties with $z\sim2$ red nuggets and are distinct from the present-day massive early-type galaxies.
As discussed in \cite{Yildirim2017}, the compact galaxies have small sizes ($\sim$$1-3$ kpc) but large stellar masses ($\sim$$10^{11}\ M_\odot$), making them consistent with the $z\sim2$ mass$-$size relation \citep{Wel2014} and strong outliers from the relation at $z\sim0$.
Likewise, their stellar mass surface densities are elevated at the center and generally show a steep fall-off at large radii, in agreement with the quiescent massive galaxy population at higher redshift.
In addition, the compact galaxies are fast rotators with disk-like shapes, consistent with the flattened structures observed for the $z\sim2$ red nuggets \citep{Wel2011}.
This attribute is in contrast to the slow rotating giant ellipticals and BCGs of today, which tend to be pressure-supported and round.

The red nuggets at $z\sim2$ are thought to form the cores of massive local elliptical galaxies, with ex-situ mass added mostly through mergers after $z\sim2$ (e.g., \citealt{Naab2009,Oser2010,Dokkum2010}).
This process would increase their bulge stellar masses and luminosities, without significantly feeding their central BHs.
Given the similarities of the local compact galaxies to the $z\sim2$ red nuggets, the compact galaxies may be passively-evolved relics of $z\sim2$ red nuggets \citep{Mateu2015,Yildirim2017}.
This idea is further supported by the uniformly old ($\sim$10 Gyr) stellar populations over the extent of the galaxies and their super-solar stellar metallicities, which are consistent with the centers of local giant ellipticals \citep{Trujillo2014, Martin2015, Mateu2017}, as well as the single populations of red globular clusters and highly concentrated dark matter halos found for individual compact galaxies \citep{Beasley2018,Buote2019}.
Cosmological hydrodynamical simulations also find that some red nuggets indeed passively evolve from $z\sim2$ to $z\sim0$ (e.g., \citealt{Wellons2016}).

While the compact galaxies could result from stripping events, the galaxies exhibit regular isophotes and do not have obvious tidal signatures in the HST images.
The compact galaxies tend to be found in cluster environments, but there are some examples of isolated field galaxies, like Mrk 1216 \citep{Yildirim2015}.
NGC 1277, NGC 1271, and UGC 2698 are all members of the Perseus cluster, but UGC 2698 is at a projected separation of $0.77^\circ$ ($\sim$1 Mpc) from NGC 1275, the BCG of the cluster, making it relatively isolated compared to NGC 1277 and NGC 1271.

If the local compact galaxies NGC 1271, NGC 1277, and Mrk 1216 are local analogs of the $z\sim2$ red nuggets, their over-massive BHs may reflect an earlier epoch when the current form of the local \mlum\ and \mmass\ relationships were not in place and galaxy growth had yet to catch up with BH growth.
However, our ALMA-based dynamical \mbh\ measurement shows that UGC 2698 does not host an over-massive BH.
We propose two plausible explanations below.

First, it is possible that the local compact galaxies are truly $z\sim2$ relics and the over-massive BHs detected so far provide evidence that the growth of BHs tends to precede the growth of stars in galaxy outskirts.
\gal\ fits on the $z\sim2$ mass-size relation and contains uniformly old stellar ages, but its status as the largest galaxy in the sample, with respect to both stellar mass and effective radius, indicates it may have experienced some stellar mass growth in its outer regions through dry mergers \citep{Yildirim2017}.
It is thus possible that \gal\ has simply experienced more stellar growth than the other local compact galaxies, increasing its luminosity and mass and bringing it more in line with the local BH scaling relations.
The other compact galaxies in the sample could be more pristine relics compared to \gal, and may thus provide better windows into the state of BH scaling relations at $z\sim2$.

Alternatively, the consistency of \gal\ with the scaling relations -- counterposed with the over-massive NGC 1271, NGC 1277, and Mrk 1216 BHs -- could indicate that there is more intrinsic scatter at the high-mass end of the \mlum\ and \mmass\ relations than currently thought.
Due to the small number of measurements, the slope and intrinsic scatter of the BH$-$galaxy correlations at the high-mass end are not well understood.
Consequently, the previously-studied compact galaxies might be less strong outliers than currently thought.
In this case, redshift evolution of the BH scaling relations is not needed to explain the over-massive BHs in the other local compact galaxies.
Going forward, more secure BH mass measurements at the upper end of the BH scaling relations, along with more \mbh\ measurements made with independent methods, will help better constrain the intrinsic scatter in the BH$-$galaxy correlations.

\section{Conclusions\label{conclusions}}

\gal\ is a member of a local compact galaxy sample consisting of plausible $z\sim2$ red nugget relics \citep{Yildirim2017}.
These quiescent galaxies have large stellar masses and velocity dispersions but are distinct from the typical present-day massive early-type galaxies.
The compact galaxies have small sizes, are fast rotators, and have cuspy surface brightness profiles.
We have used ALMA's cutting edge capabilities to study the spatial and kinematic structure of molecular gas at the center of \gal.

With 0\farcs{14}-resolution ALMA observations, we mapped spatially resolved CO(2$-$1) kinematics and fit dynamical models to derive the mass of the central BH.
From our models, we measured a BH mass of \mbhfullerr.
We explored the BH mass error budget, changing assumptions and choices made during the modeling.
We found that the statistical uncertainties are smaller than the estimated systematic uncertainties.
The dominant systematic uncertainty is associated with our treatment of dust and the subsequent description of the stellar mass distribution.
Due to the fact that we only marginally resolve the gas within the BH SOI, there remains a degeneracy between \mbh\ and \ml.
Thus, the handling of circumnuclear dust becomes important even though no significant dust attenuation at the nucleus is seen in the HST WFC3 $H$-band image.

When we place \gal\ on the BH scaling relations, we find that it is consistent with the \msig, \mlum, and \mmass\ relations.
In contrast, three other local compact galaxies from \citet{Yildirim2017} with stellar-dynamical \mbh\ measurements \citep{Walsh2015, Walsh2016, Walsh2017} are positive outliers from the \mlum\ and \mmass\ relations.
It is possible that all four nearby compact galaxies are analogs of the $z\sim2$ red nuggets, and suggest that BHs at $z\sim2$ were over-massive relative to their host galaxies.
Then, most galaxies added stellar mass to their outskirts through dry mergers \citep{Martin2015} after $z\sim2$, until they aligned with the local relations.
In this case, NGC 1271, NGC 1277, and Mrk 1216 may represent more pristine relics of the $z\sim2$ red nuggets.
\gal\ is the largest of the nearby compact galaxy sample in terms of both effective radius and mass.
The system may thus represent an intermediate evolutionary step between the $z\sim2$ red nuggets and the common massive ellipticals in the local Universe, having undergone some minor or intermediate dry mergers to bring the galaxy more in line with the local BH scaling relations.

Alternatively, there could be more intrinsic scatter in the BH$-$galaxy relations than currently thought, and the compact galaxies with over-massive BHs may not be such strong outliers after all.
In order to gain a better understanding, we need more BH measurements at the upper end of the BH mass distribution.
Continuing to measure BH masses for the local compact galaxy sample will increase the number of galaxies at the poorly sampled high-mass end of the BH relations.
Applying independent dynamical methods to a significant number of galaxies will also help better constrain the intrinsic scatter of the relations.
ALMA affords an exciting opportunity to obtain molecular gas-dynamical \mbh\ measurements for those objects in the compact galaxy sample hosting nuclear dust disks.
In a few cases, the ALMA-based inference can be directly compared to the stellar-dynamical measurement.
Such steps will help constrain whether galaxies build up their BHs before their stellar masses, or whether the intrinsic scatter at the high-mass end of the local BH scaling relations is underestimated.

\section*{Acknowledgements}
This paper makes use of the following ALMA data: ADS/JAO.ALMA\#2016.1.01010.S.
ALMA is a partnership of ESO (representing its member states), NSF (USA) and NINS (Japan), together with NRC (Canada), MOST and ASIAA (Taiwan), and KASI (Republic of Korea), in cooperation with the Republic of Chile.
The Joint ALMA Observatory is operated by ESO, AUI/NRAO and NAOJ.
The National Radio Astronomy Observatory is a facility of the National Science Foundation operated under cooperative agreement by Associated Universities, Inc.
Based on observations made with the NASA/ESA Hubble Space Telescope, obtained from the data archive at the Space Telescope Science Institute.
The observations are associated with program 13416.
STScI is operated by the Association of Universities for Research in Astronomy, Inc. under NASA contract NAS 5-26555.
Portions of this research were conducted with the advanced computing resources provided by Texas A\&M High Performance Research Computing.
J.~L.~W. was supported in part by NSF grant AST-1814799.
Research at UC Irvine was supported by NSF grant AST-1614212.
L.~C.~H. was supported by the National Science Foundation of China (11721303, 11991052) and the National Key R\&D Program of China (2016YFA0400702).
This work used the \texttt{PYTHON} Programming Language, along with community-maintained and developed software packages \texttt{ASTROPY} \citep{Astropy2013,Astropy2018}, \texttt{MATPLOTLIB} \citep{Hunter2007}, \texttt{NUMPY} \citep{Walt2011,Harris2020}, and \texttt{SCIPY} \citep{Virtanen2020}.
This work also used arXiv.org and NASA's Astrophysics Data System for bibliographic information.

\end{document}